\documentclass[aps,prl,superscriptaddress,twocolumn,floatfix]{revtex4-1}

\usepackage{amsmath}
\usepackage{amssymb}
\usepackage{amsthm}
\usepackage{graphicx}
\usepackage{graphics}
\usepackage{subfigure}
\usepackage{color}
\usepackage{bm}
\usepackage{hyperref}
\usepackage{rotating}

\renewcommand{\>}{\rangle}
\newcommand{\be}{\begin{equation} }
\newcommand{\ee}{\end{equation} }
\newcommand{\ba}{\begin{eqnarray} }
\newcommand{\ea}{\end{eqnarray} }

\newcommand{\down}{\downarrow}

\newcommand{\bpm}{\begin{pmatrix}}
\newcommand{\epm}{\end{pmatrix}}
\newcommand{\bmm}{\begin{matrix}}
\newcommand{\emm}{\end{matrix}}
\newcommand{\up}{\uparrow}

\begin{document}

\title{The effect of interactions on 2D fermionic symmetry-protected topological phases with $Z_2$ symmetry}
%\title{The effect of interactions for symmetry-protected topological phases in $2$D fermion systems}

\author{Zheng-Cheng Gu}
\affiliation{Institute for Quantum Information and Matter, California Institute
of Technology, Pasadena, CA 91125, USA}
\author{Michael Levin}
\affiliation{Condensed Matter Theory Center, Department of Physics, University of Maryland, College Park, Maryland 20742, USA}

\date{\today}

\begin{abstract}
We study the effect of interactions on $2D$ fermionic symmetry-protected topological (SPT) phases using the
recently proposed braiding statistics approach. We focus on a simple class of examples: superconductors with a $Z_2$ 
Ising symmetry. Although these systems are classified by $\mathbb{Z}$ in the noninteracting limit, our results suggest that 
the classification collapses to $\mathbb{Z}_8$ in the presence of interactions -- consistent with previous work that analyzed
the stability of the edge. Specifically, we show that there are at least $8$ different types of Ising superconductors that 
cannot be adiabatically connected to one another, even in the presence of strong interactions. In addition, we prove that each of 
the $7$ nontrivial superconductors have protected edge modes. 
\end{abstract}

%\pacs{}

\maketitle

{\it Introduction.---}
Recently it has become apparent that generalizations of topological insulators\cite{KaneMele,KaneMele2, Roy,FuKaneMele,MooreBalents,HasanKaneRMP}
known as ``symmetry-protected topological (SPT) phases''\cite{GuSPT, XieSPT1, XieSPT2,XieSPT3,XieSPT4,PollmannSPT1,PollmannSPT2,
NorbertSPT,LukaszfSPT1} can be realized in large classes of interacting boson and fermion systems. Loosely speaking, SPT phases are
characterized by two properties. First, they support robust gapless boundary modes which are protected by certain symmetries.
Second, SPT phases can be adiabatically connected to a ``trivial state''(i.e., an atomic insulator or product state) if
the relevant symmetries are broken. While significant progress has been made in understanding SPT phases in $1D$ systems,
\cite{XieSPT1,XieSPT2,PollmannSPT1,PollmannSPT2,NorbertSPT,LukaszfSPT1,LukaszfSPT2} less is known about
the higher dimensional case. Several approaches have been developed to understand these higher dimensional systems.
One approach, which applies to bosonic SPT phases in general spatial dimension, is the cohomology classification scheme of
Ref. \cite{XieSPT4,XieSPT5}. Another approach, which applies to bosonic or fermionic $2D$ SPT phases with chiral boson edge modes,
is to study the edge theories of these systems using the $K$-matrix formalism\cite{SPTCS1,SPTCS2,SPTCS3}.

In this paper, we discuss a third approach which was introduced in Ref. \onlinecite{duality} and applies to $2D$ SPT phases
with unitary symmetry groups. The key idea behind this method is to study SPT phases by ``gauging''
their symmetries -- i.e. coupling them to an appropriate gauge field, thereby transforming their global symmetries into gauge
symmetries. One can then probe the structure of the original SPT phases by constructing the excitations of the
gauged systems and computing their quasiparticle braiding statistics. This approach has several nice
features. First, it provides a simple way to distinguish different SPT phases: if two gauged systems have different quasiparticle
statistics then it is clear that the corresponding ``ungauged'' systems cannot be adiabatically connected without breaking the
symmetry. Second, it gives insight into the stability of the edge: as shown in Ref. \onlinecite{duality}, the
quasiparticle braiding statistics of the gauged system can be used to prove the existence of protected edge modes.

While Ref. \onlinecite{duality} focused on bosonic SPT phases, here we explore the fermionic case -- a problem
of particular interest because the classification of interacting fermionic SPT phases is not understood beyond $1D$ (although an
interesting attempt was made in Ref. \onlinecite{Gusuper}). We focus on
a simple class of examples: $2D$ superconductors with a $Z_2$ Ising symmetry. It was previously conjectured\cite{edge2,edge1,edge3}
that while these systems are classified by an integer invariant $\mathbb{Z}$ in the noninteracting limit, the classification 
collapses to $\mathbb{Z}_8$ when interactions are included. This claim was supported by an analysis of edge instabilities.
Here we obtain further evidence supporting this conjecture. First, we show that there are
at least $8$ different types of Ising superconductors that cannot be adiabatically connected to one another, even in the 
presence of strong interactions. Second, we prove that each of the $7$ nontrivial superconductors have protected edge modes.

{\it Pseudospin notation.---}
We begin with some notation. Consider a general fermion system with an on-site, unitary $Z_2$ symmetry $S$. Without loss of
generality, we can assume that the Hamiltonian is built out of fermion operators that have a definite
parity under $S$. \footnote{This assumption is justified since we can always choose a basis of fermion
operators in which $S$ is diagonal.} We will label the operators that are even under $S$ with a
pseudospin index $\up$ and operators that are odd under $S$ with an index $\down$.
In this notation, the system is composed out of two species of fermions, $c_\up$ and $c_\down$, where
\begin{equation}
S c_\up S^{-1}=c_\up ;\quad S c_\down S^{-1}=-c_\down
\end{equation}

In addition to the above $Z_2$ symmetry, locality dictates that the system must also conserve
fermion parity $P_f$, defined by
\begin{equation}
P_f c_\up P_f^{-1}=-c_\up ;\quad P_f c_\down P_f^{-1}=-c_\down
\end{equation}
Putting these two constraints together, we can see that the pseudospin-$\up$ and $\down$ fermions
are \emph{separately} conserved modulo $2$.
%Thus the global symmetry of the system is $Z_2^\up \otimes Z_2^\down$.

{\it The noninteracting limit.---}
We next review the classification of \emph{noninteracting} fermion SPT phases with $Z_2$ Ising symmetry.
The key observation is that quadratic pseudospin mixing terms, e.g. $c^\dagger_\up c_\down$, are prohibited by the $Z_2$
symmetry. Therefore the $c_\up$ and $c_\down$ fermions are completely decoupled in the non-interacting limit.
Applying the known integer classification of 2D topological superconductors\cite{RyuSPT,Kitaevperiod}, it follows that
the different free fermion phases are classified by a pair of integers $(\nu^\up,\nu^\down)$.
Here, $(\nu^\up,\nu^\down)\in \mathbb{Z}^2$ corresponds to a phase where the pseudospin-$\up$ and pseudospin-$\down$
fermions form two decoupled topological superconductors with $\nu^\up$ and $\nu^\down$ chiral Majorana edge modes, respectively.
(The sign of $\nu^\up, \nu^\down$ indicates the chirality of the edge mode -- left or right moving).

In this paper, we only consider a \emph{subset} of the above phases -- namely those satisfying
$\nu^\up = -\nu^\down$. The reason for this restriction is that our definition for SPT phases requires
that they be adiabatically connected to a trivial band insulator if the symmetry is
broken, and only phases with $\nu^\up = -\nu^\down$ obey this condition. Hence, according to our definition,
the noninteracting SPT phases are classified by a single integer $\nu = \nu^\up = -\nu^\down$.

%\begin{figure}[t]
%\begin{center}
%\vskip -0.5cm
%\includegraphics[width=6cm]{free.pdf}
%\vskip -0.5cm
%\caption{(color online)
%A free fermion SPT phase protected by an $Z_2$(Ising) symmetry can be regarded as two decoupled layers of free fermions with opposite
%Chern number. The fermion parity is conserved for each individual layer. By coupling to two dynamic $Z_2$ gauge field associated with the two %individual fermion parity conservation for each layer, we map free fermion
%SPT phases to intrinsic topologically ordered phases. Different intrinsic topologically ordered phase will correspond to a different SPT phase.
%}\label{fig:free}
%\end{center}
%\vskip -1.0cm
%\end{figure}

{\it The effect of interactions.---}
While the $Z_2$ symmetry requires that the pseudospin $\up$ and $\down$ fermions decouple
from one another in the non-interacting limit, interspecies coupling is allowed once we add interactions into the system.
(For example, the four fermion term $c_{i\up}^\dagger c_{j\up}^\dagger c_{j\down} c_{i\down}$ is $Z_2$ symmetric,
but mixes the two species). Thus, we might expect that the $\mathbb{Z}$ classification will collapse once we
include interactions: i.e. it may be possible to adiabatically connect phases with different values of $\nu$. The question we
will now investigate is: how many distinct phases survive in the presence of interactions?

For concreteness, we focus our analysis on a free fermion Hamiltonian with nearest neighbor(NN) hopping and pairing
terms:
\begin{eqnarray}
H^\nu &=&\sum_{\lambda=1 }^\nu \sum_\sigma\sum_{\langle ij\rangle}t_{ij\sigma}c_{i\sigma;\lambda}^\dagger c_{j\sigma;\lambda}+h.c.
 \nonumber\\
 &+&\sum_{\lambda=1 }^\nu \sum_\sigma\sum_{\langle ij\rangle}t_{ij\sigma}\Delta_{ij\sigma}c_{i\sigma;\lambda}^\dagger c_{j\sigma;\lambda}^\dagger+h.c.\label{Hsimple}
\end{eqnarray}
Here $i$ runs over lattice sites, while $\lambda = 1,..., \nu$ describes different orbital states on each lattice site.
We choose the the pairing term $\Delta_{ij;\up(\down)}$ so as to describe a $p+ip \ (p-ip)$ superconductor, with an edge containing
a single right (left) moving chiral Majorana mode.
Our task is to determine which $H^\nu$ describe distinct phases, and which can be adiabatically connected
to one another. Our strategy for answering this question is to couple the pseudospin $\up$
and $\down$ fermions to two independent $Z_2$ gauge fields, ($Z_2^\up \times Z_2^\down$) and then study the braiding
statistics of the $Z_2$ flux excitations in the gauged model. We will show that some values of $\nu$
exhibit different braiding statistics and therefore must represent distinct phases.

The gauged model that we will analyze can be formally written as
\begin{eqnarray}
&&H^{\rm{gauge}}=\sum_{\lambda=1 }^\nu\sum_\sigma\sum_{\langle ij\rangle}
t_{ij\sigma}c_{i\sigma;\lambda}^\dagger \tau_{ij\sigma}^z c_{j\sigma;\lambda}+h.c. \nonumber\\
&+&\sum_{\lambda=1 }^\nu\sum_\sigma\sum_{\langle ij\rangle}\Delta_{ij\sigma}c_{i\sigma;\lambda}^\dagger
\tau_{ij\sigma}^z c_{j\sigma;\lambda}^\dagger+h.c.-H_\sigma^{\rm{flux}}
\label{gaugetotal}
\end{eqnarray}
where $\tau_{ij\sigma}^z$ is the $Z_2$ gauge field strength associated with pseudospin $\sigma$ fermions
and where $H_\sigma^{\rm{flux}} = \sum_{\<ijkl\>}\tau_{ij \sigma}^z \tau_{jk \sigma}^z \tau_{kl \sigma}^z \tau_{li \sigma}^z$
is a flux energy term that gives an energy cost to flux excitations of the gauge fields.
($H_\sigma^{\rm{flux}}$ is the analogue of the $\mathbf{B}^2$ term in Maxwell electromagnetic dynamics). The
Hamiltonian $H^{\rm{gauge}}$ is defined in a Hilbert space consisting of gauge invariant states -- that
is, all states satisfying the constraint $\prod _{j\in NN(i)} \tau_{ij\sigma}^x=(-)^{\sum_\lambda n_{i\sigma;\lambda}}$.
This constraint can be thought of as a $Z_2$ analogue of Gauss' law, $\nabla \cdot \textbf{E}=\rho$.

The next step is to compare the quasiparticle braiding statistics of the gauged
system Eq. (\ref{gaugetotal}) for different values of $\nu$. To this end, it is useful to first think about
a simpler system with only one pseudospin component and $\nu$ chiral edge modes. The quasiparticle braiding
statistics of such a chiral superconductor were worked out by Kitaev in Ref. \onlinecite{KitaevSC}. That calculation
showed that the quasiparticle braiding statistics of the superconductor depends on the number
of chiral edge modes $\nu$, modulo $16$. For example, if $\nu$ is even, the $Z_2$ gauge fluxes (i.e. superconducting vortices)
are Abelian anyons with an exchange phase factor $e^{\frac{\pi}{8}i\nu}$. If $\nu$ is odd, the $Z_2$-fluxes are
non-Abelian anyons with an exchange phase $(-)^{\frac{(\nu^2-1)}{8}}e^{\frac{\pi}{8}i\nu}$ when the two
non-Abelian anyons are in the vacuum fusion channel.

Now, let us consider the full system, which consists of a pseudospin $\up$ component with $\nu$ right
moving edge modes and a pseudospin $\down$ component with $\nu$ left moving edge modes. Naively, one
might guess that the braiding statistics of this system also depends on $\nu$ modulo $16$, since it
is made up of two independent chiral superconductors. However, this guess is incorrect: the
braiding statistics of the ``doubled'' system only depends on $\nu$ modulo $8$. To see this, we
need to show that the braiding statistics for $\nu=0,1,...,7$ are all different while the $\nu=0$ case is equivalent
to the $\nu=8$ case. One way to establish the first statement is to compute the exchange phases of all
the different types of $Z_2^\down$ (or $Z_2^\up$) flux excitations. Here, a $Z_2^\down$ flux is defined to
be any quasiparticle excitation that acquires a phase of $-1$ when braided around a pseudospin-$\down$ fermion
and acquires no phase when braided around a pseudospin-$\up$ fermion. Using the results of
Ref. \onlinecite{KitaevSC}, it is easy to see that for even $\nu$ there are $4$ types
of $Z_2^\down$ fluxes with exchange statistics $\pm e^{\frac{\pi}{8}i\nu}$, while for odd $\nu$ there are
$2$ types of $Z_2^\down$ fluxes with exchange statistics $\pm e^{-\frac{\pi}{8}i\nu}$. (In the latter case, we assume
the fluxes are in the vacuum fusion channel). In particular, we see that
the exchange statistics of the $Z_2$ fluxes are different for each of the eight possibilities $\nu = 0,1,...,7$.

On the other hand, to see that $\nu =0$ and $\nu = 8$ have the same braiding statistics, we need to construct
an explicit isomorphism between the quasiparticles in the two systems. To this end, we consult
Ref. \onlinecite{KitaevSC} and note that for both $\nu = 0,8$ the gauge theory Eq.(\ref{gaugetotal}) has four quasiparticles
$1,e_\sigma,m_\sigma,\varepsilon_\sigma$ for each pseudospin direction, $\sigma = \up, \down$. Including all possible composites of pseudospin $\up$ and $\down$
excitations, there are $4 \cdot 4 = 16$ quasiparticles all together. We can think of the $\varepsilon_\sigma$
as the constituent fermions while $e_\sigma$ and $m_\sigma$ are different types of $Z^\sigma_2$ gauge
fluxes which differ from one another by the addition of a fermion: $e_\sigma = m_\sigma \cdot \varepsilon_\sigma$.
Using the results of Ref. \onlinecite{KitaevSC}, we can see that for both $\nu = 0,8$, the three particles
$\varepsilon_\sigma, e_\sigma, m_\sigma$ acquire a phase of $-1$ when braided around each other. The only difference
is that $e_\sigma$ and $m_\sigma$ are \emph{bosons} for the case $\nu=0$ while they are \emph{fermions} for the case $\nu=8$. With these
properties in mind, one can easily see that the following map gives an isomorphism between the quasiparticles in the
two systems:
\begin{widetext}
\begin{table}[h]
\begin{tabular}{|c||c|c|c|c|c|c|c|c|c|c||c|c|c|c|c|c|}
\hline
$\nu=0$ & $1$ & $e_\up$ & $m_\up$  & $e_\down$ & $m_\down$ & $e_\up e_\down$
& $m_\up m_\down$ & $\varepsilon_\up\varepsilon_\down$ & $e_\up m_\down$ & $m_\up e_\down$
& $\varepsilon_\up$ & $\varepsilon_\down$ & $e_\up\varepsilon_\down$ & $\varepsilon_\up e_\down$ & $m_\up\varepsilon_\down$
& $\varepsilon_\up m_\down$ \\

\hline
$\nu=8$ & $1$ & $e_\up\varepsilon_\down$ & $m_\up\varepsilon_\down$ & $\varepsilon_\up e_\down$
& $\varepsilon_\up m_\down$ & $m_\up m_\down$ & $e_\up e_\down$ & $\varepsilon_\up\varepsilon_\down$
& $m_\up e_\down$ & $e_\up m_\down$ & $\varepsilon_\up$ & $\varepsilon_\down$
& $e_\up$ & $e_\down$ & $m_\up$ & $m_\down$  \\
\hline
\end{tabular}
\end{table}
\end{widetext}
Here, the table is organized so that the first ten quasiparticles are all bosons while the other six are all fermions.
We can see that the correspondence not only preserves braiding statistics and fusion rules, but also preserves the
$Z^\up_2 \times Z^\down_2$ gauge structure, mapping the $\down$ fermions ($\varepsilon_\down$)
of one system onto the corresponding fermions in the other system, and likewise mapping the $Z^\down_2$ fluxes
($e_\down, m_\down, \varepsilon_\up e_\down, \varepsilon_\up m_\down$) of one system onto the $Z^\down_2$ fluxes of the other system
(and similarly for $\up$).

Two conclusions follow from the above analysis. First, we conclude that the Hamiltonians $H^\nu$ with $\nu = 0,1,\cdots,7$
cannot be adiabatically connected to one another without breaking the $Z_2$ symmetry. Indeed, if there existed a gapped
path connecting these Hamiltonians, then there would have to be a corresponding path connecting the
gauged systems $H^{\rm{gauge}}$ -- an impossibility, since we have seen that they have different quasiparticle braiding statistics.
The second conclusion is that it is at least \emph{plausible} that $H^0$ and $H^8$ can be adiabatically connected to one another
in the presence of interactions, since the corresponding $Z_2^\up \times Z_2^\down$ gauge theories share the same
statistics and gauge structure.

{\it The instability of $\nu=8$ edge.---}  In this section, we give additional evidence that the $\nu = 8$ system is a trivial SPT
phase: we show that the $\nu = 8$ edge can be gapped out by appropriate interactions, without breaking the $Z_2$ symmetry
(explicitly or spontaneously). We note that a similar result was obtained previously in 
Refs. \onlinecite{edge1,edge2,edge3}.

Our approach is based on bosonization. We note that the edge of the $\nu = 8$ free fermion Hamiltonian
Eq.(\ref{Hsimple}) contains $8$ pseudospin-$\up$ Majorana modes and $8$ pseudospin-$\down$ Majorana modes
moving in opposite directions. Pairing up the Majorana modes to form complex fermions, we can
equivalently describe the edge using $4$ pseudospin-$\up$ and $4$ pseudospin-$\down$ complex fermions.
We then bosonize these fermions, using $4$ boson modes $\Phi_1,..,\Phi_4$ for
the pseudospin-$\up$ fermions, and $4$ boson modes $\Phi_5,...,\Phi_8$ for the pseudospin-$\down$ fermions.
The edge is then described by the chiral boson Lagrangian
\begin{equation}
\mathcal{L}_{edge}= \frac{1}{4\pi} (K_{IJ} \partial_x\Phi_I \partial_x\Phi_J - V_{IJ} \partial_x \Phi_I \partial_x\Phi_J)
\end{equation}
where $K=diag(1,1,1,1,-1,-1,-1,-1)$, and $V_{IJ}$ is the velocity matrix. Here we use a
normalization convention where the fermion creation operators are of the form $e^{i\Phi_k}$, $k=1,...,8$.
In this language, the symmetry transformation is given by
$S^{-1} \Phi S = \Phi + \pi K^{-1} \chi$
where $\chi^T = (0,0,0,0,1,1,1,1)$.

We now construct interaction terms that gap out the edge without breaking
the $Z_2$ symmetry (either explicitly or spontaneously). We consider backscattering terms of the form
$U(\Lambda) = U(x) \cos(\Lambda^T K \Phi - \alpha(x))$.
In order for $U(\Lambda)$ to be invariant under $S$, we require that
\begin{equation}
\Lambda^T \chi \equiv 0 \text{ (mod $2$)}
\label{symmcond}
\end{equation}

In order to gap out the edge, we need to add $4$ backscattering terms $\sum_i U(\Lambda_i)$: each term can gap out a pair of
counter-propagating edge modes. Such terms can gap out the edge as long as the $\{\Lambda_i\}$ vectors satisfy \cite{Haldane}
\begin{equation}
\Lambda_i^T K \Lambda_j = 0
\label{nullcond}
\end{equation}
for all $i,j$. This ``null-vector'' condition guarantees that we can make a suitable change
of variables mapping $L_{edge}$ onto a system of $4$ decoupled Luttinger liquids with $4$ backscattering terms.
It is then easy to see that the backscattering terms will gap out the corresponding Luttinger liquids (at least for large $U$
\footnote{Note that it is not important to us whether these terms are relevant or irrelevant in the renormalization group sense. The
reason is that we are not interested in the \emph{perturbative} stability of the edge, but rather whether it is stable to arbitrary
perturbations that do not break the symmetry, explicitly or spontaneously.}).

We now claim that the following $\{\Lambda_i\}$ will do the job:
\begin{eqnarray}
\Lambda_1^T &=& (1,-1, 0, 0, 1, -1, 0, 0); \nonumber\\
\Lambda_2^T &=& (1, 0, -1, 0, 1, 0, -1,0);\nonumber\\
\Lambda_3^T &=& (1, 0, 0, -1, 1, 0, 0, -1); \nonumber\\
\Lambda_4^T &=& (1, 0, 1, 0, 0, -1, 0, -1).
\end{eqnarray}
Indeed, it is easy to check that these $\{\Lambda_i\}$ obey the null vector criterion (\ref{nullcond}), as well as
the symmetry condition (\ref{symmcond}). To complete the argument, we need to check that the perturbation corresponding
to $\{\Lambda_i\}$ does not spontaneously break the $Z_2$ symmetry. However, as explained in Ref. \cite{SPTCS1}, we can rule out
the possibility of spontaneous symmetry breaking if the $\bpm 8 \\ 4 \epm$ $4 \times 4$ minors of the
$8 \times 4$ matrix with columns $\Lambda_1,...,\Lambda_4$ have no common factor. This property of $\Lambda_1,...,\Lambda_4$
can be verified by direct calculation.

{\it Protected edge states for $\nu\neq 0 \text{ mod }8$ .---}
On the other hand, we now show that the edge of $H^\nu$ is \emph{protected} if $\nu\neq 0 \text{ mod }8$. 
To state our result more precisely, let us consider a disk geometry and a Hamiltonian of the form
$H = H_{\text{bulk}} + H_{\text{edge}}$, where $H_{\text{bulk}} = H^\nu$, and $H_{\text{edge}}$ is an arbitrary
interacting Hamiltonian acting on fermions near the edge. In this setup, what we will show is that the ground
state $|0\>$ cannot be both $Z_2$ symmetric and ``short-range entangled.'' \footnote{Here, a fermionic state is said to be
``short-range entangled'' if it can be transformed into an atomic insulator by a local unitary transformation --
a unitary operator generated from the time evolution of a local Hamiltonian over a finite time $t$.} We believe that
this result rules out the possibility of a $Z_2$ symmetric, gapped edge, and in this sense proves that the gapless
edge excitations are protected.

\begin{figure}[tb]
\centerline{
\includegraphics[width=0.8\columnwidth]{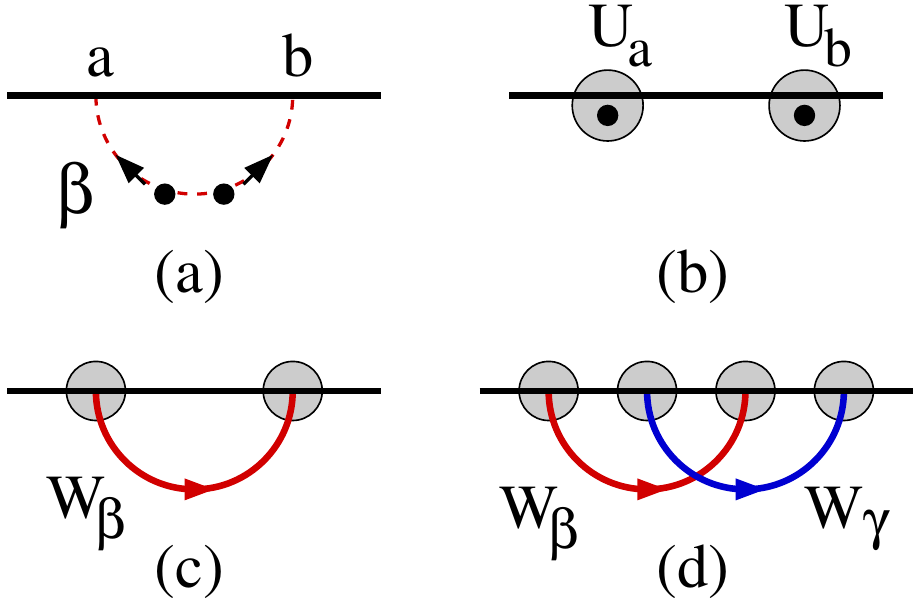}
}
\caption{
(a) We consider a thought experiment in which we create two $Z^\down_2$ fluxes in the bulk and then move them
along a path $\beta$ to points $a,b$ at the edge.
(b) We argue that the two fluxes can be annihilated at the boundary by applying local operators $U_a$, $U_b$.
(c) We define $\mathbb{W}_\beta$ to be an operator which describes a process in which the fluxes are created
in the bulk, brought to the edge, and then annihilated.
(d) To obtain a contradiction, we consider two paths $\beta, \gamma$ that intersect one another, and we investigate
the commutation algebra of the corresponding operators $\mathbb{W}_\beta$,$\mathbb{W}_\gamma$.
}
\vskip -0.5cm
\label{braidfig}
\end{figure}

As in Ref. \cite{duality}, our argument is a proof by contradiction: we assume that $|0\>$
is short-range entangled and $Z_2$ symmetric and we show that these assumptions lead to a contradiction.
The first step is to couple the pseudospin-$\up$ and pseudospin-$\down$ fermions to two independent
$Z_2$ gauge fields as in (\ref{gaugetotal}). We then imagine creating a pair of $Z^{\down}_2$ (or $Z^{\up}_2$)
fluxes in the bulk. After creating the $Z^{\down}_2$ fluxes, we separate them and move them along
some path $\beta$ to points $a,b$ at the boundary (Fig. \ref{braidfig}a). Formally, this process can be
implemented by applying a unitary (string-like) operator $W_\beta$ to $|0\>$.

Next, we claim that the $Z^\down_2$ fluxes can be annihilated at the boundary if we apply appropriate local
operators. That is, there exist local operators $U_a, U_b$, acting near points $a,b$ such that
$U_a U_b W_\beta |0\> = |0\>$ (Fig. \ref{braidfig}b). Establishing this claim is the hardest step in
the argument, and here we merely outline its proof.\footnote{For a more detailed proof in a related bosonic system,
see Ref. \onlinecite{duality}.} The basic point is that when we bring the $Z^\down_2$ fluxes to the boundary,
we effectively create two $Z^\down_2$ domain walls at $a$ and $b$. Given that the ground state is $Z^\down_2$
symmetric, these domain walls are \emph{local} excitations: they only affect expectation values in the neighborhood
of $a$ and $b$. It then follows that these domain walls can be annihilated by local
operators since local excitations of a short-range entangled state can always be annihilated locally.

In the third step, we consider a creation and annihilation process in which two $Z^\down_2$ fluxes are created in
the bulk, moved to the boundary and then annihilated. Let $\mathbb{W}_\beta$ be a unitary operator describing this
process (Fig. \ref{braidfig}c). (Formally, $\mathbb{W}_\beta = U_a U_b W_\beta$). Now, consider a second path
$\gamma$ with the geometry shown in Fig. \ref{braidfig}d and define $\mathbb{W}_\gamma$ in the same way. By
construction, we have $\mathbb{W}_\beta |0\> = \mathbb{W}_\gamma |0\> = |0\>$. Hence
\begin{equation}
\mathbb{W}_\beta \mathbb{W}_\gamma |0\> = \mathbb{W}_\gamma \mathbb{W}_\beta |0\> = |0\>
\label{WW1}
\end{equation}

In the final step, we show that (\ref{WW1}) leads to a contradiction if $\nu \neq 0 \text{ (mod $8$)}$.
It is useful to consider separately the case where $\nu$ is even and $\nu$ is odd. First, suppose $\nu$ is even.
In this case, the  $Z^\down_2$ fluxes are abelian anyons and it follows from a general analysis
of abelian quasiparticle statistics (see e.g. Refs. \onlinecite{duality,LevinWenHop}) that
\begin{equation}
\mathbb{W}_\beta \mathbb{W}_\gamma |0\> = e^{2i\theta} \mathbb{W}_\gamma \mathbb{W}_\beta |0\>
\label{WW2}
\end{equation}
where $e^{i\theta}$ is the exchange phase of the $Z_2^\down$ fluxes. According to the braiding statistics
calculation outlined above, the four types of $Z^\down_2$ fluxes have exchange statistics
$\theta= \pm \frac{\pi \nu}{8}$. Hence if $\nu \neq 0 \text{ (mod $8$)}$ then $e^{2i\theta} \neq 1$ for any of the
four types of fluxes and equations (\ref{WW1}), (\ref{WW2}) are in contradiction.

Now suppose $\nu$ is odd. In this case, the $Z_2^\down$ fluxes are non-abelian anyons, so the above braiding statistics analysis
is more complicated. However, we can avoid these complications using an alternative argument. We note that if $\nu$ is odd then
each $Z_2^\down$ flux carries an \emph{unpaired Majorana mode}. Thus, the state $W_\beta |0\>$ has unpaired Majorana modes
localized near points $a$ and $b$. But then it is clearly impossible for $U_a U_b W_\beta |0\> = |0\>$ since unpaired Majorana
modes cannot be destroyed by any local operation. Once again, we encounter a contradiction, implying that our assumption
is false and $|0\>$ cannot be both $Z_2$ symmetric and short-range entangled.

{\it Conclusion.---}
In this paper we have studied SPT phases in interacting fermion systems using a braiding statistics approach. As a simple
example, we considered superconductors with a $Z_2$ (Ising) symmetry. Although in the noninteracting case these 
Ising superconductors are classified by an integer
invariant $\nu \in \mathbb{Z}$, we give evidence that the classification collapses to $\mathbb{Z}_8$ in the presence
of interactions. We also give a general argument proving that the edge excitations are protected when $\nu \neq 0 \text{ (mod $8$)}$
and unprotected when $\nu = 0 \text{ (mod $8$)}$.

{\it Acknowledgement.---}
ZCG is supported in part by NSF Frontiers Center with support from the Gordon and Betty Moore Foundation.
ML acknowledges support from the Alfred P. Sloan foundation.
\bibliography{dualityfermion}

\begin{thebibliography}{34}%
\makeatletter
\providecommand \@ifxundefined [1]{%
 \@ifx{#1\undefined}
}%
\providecommand \@ifnum [1]{%
 \ifnum #1\expandafter \@firstoftwo
 \else \expandafter \@secondoftwo
 \fi
}%
\providecommand \@ifx [1]{%
 \ifx #1\expandafter \@firstoftwo
 \else \expandafter \@secondoftwo
 \fi
}%
\providecommand \natexlab [1]{#1}%
\providecommand \enquote  [1]{``#1''}%
\providecommand \bibnamefont  [1]{#1}%
\providecommand \bibfnamefont [1]{#1}%
\providecommand \citenamefont [1]{#1}%
\providecommand \href@noop [0]{\@secondoftwo}%
\providecommand \href [0]{\begingroup \@sanitize@url \@href}%
\providecommand \@href[1]{\@@startlink{#1}\@@href}%
\providecommand \@@href[1]{\endgroup#1\@@endlink}%
\providecommand \@sanitize@url [0]{\catcode `\\12\catcode `\$12\catcode
  `\&12\catcode `\#12\catcode `\^12\catcode `\_12\catcode `\%12\relax}%
\providecommand \@@startlink[1]{}%
\providecommand \@@endlink[0]{}%
\providecommand \url  [0]{\begingroup\@sanitize@url \@url }%
\providecommand \@url [1]{\endgroup\@href {#1}{\urlprefix }}%
\providecommand \urlprefix  [0]{URL }%
\providecommand \Eprint [0]{\href }%
\providecommand \doibase [0]{http://dx.doi.org/}%
\providecommand \selectlanguage [0]{\@gobble}%
\providecommand \bibinfo  [0]{\@secondoftwo}%
\providecommand \bibfield  [0]{\@secondoftwo}%
\providecommand \translation [1]{[#1]}%
\providecommand \BibitemOpen [0]{}%
\providecommand \bibitemStop [0]{}%
\providecommand \bibitemNoStop [0]{.\EOS\space}%
\providecommand \EOS [0]{\spacefactor3000\relax}%
\providecommand \BibitemShut  [1]{\csname bibitem#1\endcsname}%
\let\auto@bib@innerbib\@empty
%</preamble>
\bibitem [{\citenamefont {Kane}\ and\ \citenamefont
  {Mele}(2005{\natexlab{a}})}]{KaneMele}%
  \BibitemOpen
  \bibfield  {author} {\bibinfo {author} {\bibfnamefont {C.~L.}\ \bibnamefont
  {Kane}}\ and\ \bibinfo {author} {\bibfnamefont {E.~J.}\ \bibnamefont
  {Mele}},\ }\href {\doibase 10.1103/PhysRevLett.95.226801} {\bibfield
  {journal} {\bibinfo  {journal} {Phys. Rev. Lett.}\ }\textbf {\bibinfo
  {volume} {95}},\ \bibinfo {pages} {226801} (\bibinfo {year}
  {2005}{\natexlab{a}})}\BibitemShut {NoStop}%
\bibitem [{\citenamefont {Kane}\ and\ \citenamefont
  {Mele}(2005{\natexlab{b}})}]{KaneMele2}%
  \BibitemOpen
  \bibfield  {author} {\bibinfo {author} {\bibfnamefont {C.~L.}\ \bibnamefont
  {Kane}}\ and\ \bibinfo {author} {\bibfnamefont {E.~J.}\ \bibnamefont
  {Mele}},\ }\href {\doibase 10.1103/PhysRevLett.95.226801} {\bibfield
  {journal} {\bibinfo  {journal} {Phys. Rev. Lett.}\ }\textbf {\bibinfo
  {volume} {95}},\ \bibinfo {pages} {146802} (\bibinfo {year}
  {2005}{\natexlab{b}})}\BibitemShut {NoStop}%
\bibitem [{\citenamefont {Roy}(2009)}]{Roy}%
  \BibitemOpen
  \bibfield  {author} {\bibinfo {author} {\bibfnamefont {R.}~\bibnamefont
  {Roy}},\ }\href {\doibase 10.1103/PhysRevB.79.195322} {\bibfield  {journal}
  {\bibinfo  {journal} {Phys. Rev. B}\ }\textbf {\bibinfo {volume} {79}},\
  \bibinfo {pages} {195322} (\bibinfo {year} {2009})}\BibitemShut {NoStop}%
\bibitem [{\citenamefont {Fu}\ \emph {et~al.}(2007)\citenamefont {Fu},
  \citenamefont {Kane},\ and\ \citenamefont {Mele}}]{FuKaneMele}%
  \BibitemOpen
  \bibfield  {author} {\bibinfo {author} {\bibfnamefont {L.}~\bibnamefont
  {Fu}}, \bibinfo {author} {\bibfnamefont {C.~L.}\ \bibnamefont {Kane}}, \ and\
  \bibinfo {author} {\bibfnamefont {E.~J.}\ \bibnamefont {Mele}},\ }\href
  {\doibase 10.1103/PhysRevLett.98.106803} {\bibfield  {journal} {\bibinfo
  {journal} {Phys. Rev. Lett.}\ }\textbf {\bibinfo {volume} {98}},\ \bibinfo
  {pages} {106803} (\bibinfo {year} {2007})}\BibitemShut {NoStop}%
\bibitem [{\citenamefont {Moore}\ and\ \citenamefont
  {Balents}(2007)}]{MooreBalents}%
  \BibitemOpen
  \bibfield  {author} {\bibinfo {author} {\bibfnamefont {J.~E.}\ \bibnamefont
  {Moore}}\ and\ \bibinfo {author} {\bibfnamefont {L.}~\bibnamefont
  {Balents}},\ }\href {\doibase 10.1103/PhysRevB.75.121306} {\bibfield
  {journal} {\bibinfo  {journal} {Phys. Rev. B}\ }\textbf {\bibinfo {volume}
  {75}},\ \bibinfo {pages} {121306} (\bibinfo {year} {2007})}\BibitemShut
  {NoStop}%
\bibitem [{\citenamefont {Hasan}\ and\ \citenamefont
  {Kane}(2010)}]{HasanKaneRMP}%
  \BibitemOpen
  \bibfield  {author} {\bibinfo {author} {\bibfnamefont {M.~Z.}\ \bibnamefont
  {Hasan}}\ and\ \bibinfo {author} {\bibfnamefont {C.~L.}\ \bibnamefont
  {Kane}},\ }\href@noop {} {\bibfield  {journal} {\bibinfo  {journal} {Rev.
  Mod. Phys.}\ }\textbf {\bibinfo {volume} {82}},\ \bibinfo {pages} {3045}
  (\bibinfo {year} {2010})}\BibitemShut {NoStop}%
\bibitem [{\citenamefont {Gu}\ and\ \citenamefont {Wen}(2009)}]{GuSPT}%
  \BibitemOpen
  \bibfield  {author} {\bibinfo {author} {\bibfnamefont {Z.-C.}\ \bibnamefont
  {Gu}}\ and\ \bibinfo {author} {\bibfnamefont {X.-G.}\ \bibnamefont {Wen}},\
  }\href {\doibase 10.1103/PhysRevB.80.155131} {\bibfield  {journal} {\bibinfo
  {journal} {Phys. Rev. B}\ }\textbf {\bibinfo {volume} {80}},\ \bibinfo
  {pages} {155131} (\bibinfo {year} {2009})}\BibitemShut {NoStop}%
\bibitem [{\citenamefont {Chen}\ \emph
  {et~al.}(2011{\natexlab{a}})\citenamefont {Chen}, \citenamefont {Gu},\ and\
  \citenamefont {Wen}}]{XieSPT1}%
  \BibitemOpen
  \bibfield  {author} {\bibinfo {author} {\bibfnamefont {X.}~\bibnamefont
  {Chen}}, \bibinfo {author} {\bibfnamefont {Z.-C.}\ \bibnamefont {Gu}}, \ and\
  \bibinfo {author} {\bibfnamefont {X.-G.}\ \bibnamefont {Wen}},\ }\href
  {\doibase 10.1103/PhysRevB.83.035107} {\bibfield  {journal} {\bibinfo
  {journal} {Phys. Rev. B}\ }\textbf {\bibinfo {volume} {83}},\ \bibinfo
  {pages} {035107} (\bibinfo {year} {2011}{\natexlab{a}})}\BibitemShut
  {NoStop}%
\bibitem [{\citenamefont {Chen}\ \emph
  {et~al.}(2011{\natexlab{b}})\citenamefont {Chen}, \citenamefont {Gu},\ and\
  \citenamefont {Wen}}]{XieSPT2}%
  \BibitemOpen
  \bibfield  {author} {\bibinfo {author} {\bibfnamefont {X.}~\bibnamefont
  {Chen}}, \bibinfo {author} {\bibfnamefont {Z.-C.}\ \bibnamefont {Gu}}, \ and\
  \bibinfo {author} {\bibfnamefont {X.-G.}\ \bibnamefont {Wen}},\ }\href
  {\doibase 10.1103/PhysRevB.84.235128} {\bibfield  {journal} {\bibinfo
  {journal} {Phys. Rev. B}\ }\textbf {\bibinfo {volume} {84}},\ \bibinfo
  {pages} {235128} (\bibinfo {year} {2011}{\natexlab{b}})}\BibitemShut
  {NoStop}%
\bibitem [{\citenamefont {Chen}\ \emph
  {et~al.}(2011{\natexlab{c}})\citenamefont {Chen}, \citenamefont {Liu},\ and\
  \citenamefont {Wen}}]{XieSPT3}%
  \BibitemOpen
  \bibfield  {author} {\bibinfo {author} {\bibfnamefont {X.}~\bibnamefont
  {Chen}}, \bibinfo {author} {\bibfnamefont {Z.-X.}\ \bibnamefont {Liu}}, \
  and\ \bibinfo {author} {\bibfnamefont {X.-G.}\ \bibnamefont {Wen}},\ }\href
  {\doibase 10.1103/PhysRevB.84.235141} {\bibfield  {journal} {\bibinfo
  {journal} {Phys. Rev. B}\ }\textbf {\bibinfo {volume} {84}},\ \bibinfo
  {pages} {235141} (\bibinfo {year} {2011}{\natexlab{c}})}\BibitemShut
  {NoStop}%
\bibitem [{\citenamefont {Chen}\ \emph {et~al.}(2013)\citenamefont {Chen},
  \citenamefont {Gu}, \citenamefont {Liu},\ and\ \citenamefont
  {Wen}}]{XieSPT4}%
  \BibitemOpen
  \bibfield  {author} {\bibinfo {author} {\bibfnamefont {X.}~\bibnamefont
  {Chen}}, \bibinfo {author} {\bibfnamefont {Z.-C.}\ \bibnamefont {Gu}},
  \bibinfo {author} {\bibfnamefont {Z.-X.}\ \bibnamefont {Liu}}, \ and\
  \bibinfo {author} {\bibfnamefont {X.-G.}\ \bibnamefont {Wen}},\ }\href
  {\doibase 10.1103/PhysRevB.87.155114} {\bibfield  {journal} {\bibinfo
  {journal} {Phys. Rev. B}\ }\textbf {\bibinfo {volume} {87}},\ \bibinfo
  {pages} {155114} (\bibinfo {year} {2013})}\BibitemShut {NoStop}%
\bibitem [{\citenamefont {Pollmann}\ \emph {et~al.}(2009)\citenamefont
  {Pollmann}, \citenamefont {Berg}, \citenamefont {Turner},\ and\ \citenamefont
  {Oshikawa}}]{PollmannSPT1}%
  \BibitemOpen
  \bibfield  {author} {\bibinfo {author} {\bibfnamefont {F.}~\bibnamefont
  {Pollmann}}, \bibinfo {author} {\bibfnamefont {E.}~\bibnamefont {Berg}},
  \bibinfo {author} {\bibfnamefont {A.~M.}\ \bibnamefont {Turner}}, \ and\
  \bibinfo {author} {\bibfnamefont {M.}~\bibnamefont {Oshikawa}},\ }\href@noop
  {} {\bibfield  {journal} {\bibinfo  {journal} {arXiv:0909.4059}\ } (\bibinfo
  {year} {2009})}\BibitemShut {NoStop}%
\bibitem [{\citenamefont {Pollmann}\ \emph {et~al.}(2010)\citenamefont
  {Pollmann}, \citenamefont {Tuner}, \citenamefont {Berg},\ and\ \citenamefont
  {Oshikawa}}]{PollmannSPT2}%
  \BibitemOpen
  \bibfield  {author} {\bibinfo {author} {\bibfnamefont {F.}~\bibnamefont
  {Pollmann}}, \bibinfo {author} {\bibfnamefont {A.~M.}\ \bibnamefont {Tuner}},
  \bibinfo {author} {\bibfnamefont {e.}~\bibnamefont {Berg}}, \ and\ \bibinfo
  {author} {\bibfnamefont {M.}~\bibnamefont {Oshikawa}},\ }\href {\doibase
  10.1103/PhysRevB.81.064439} {\bibfield  {journal} {\bibinfo  {journal} {Phys.
  Rev. B}\ }\textbf {\bibinfo {volume} {81}},\ \bibinfo {pages} {064439}
  (\bibinfo {year} {2010})}\BibitemShut {NoStop}%
\bibitem [{\citenamefont {Schuch}\ \emph {et~al.}(2011)\citenamefont {Schuch},
  \citenamefont {Perez-Garcia},\ and\ \citenamefont {Cirac}}]{NorbertSPT}%
  \BibitemOpen
  \bibfield  {author} {\bibinfo {author} {\bibfnamefont {N.}~\bibnamefont
  {Schuch}}, \bibinfo {author} {\bibfnamefont {D.}~\bibnamefont
  {Perez-Garcia}}, \ and\ \bibinfo {author} {\bibfnamefont {I.}~\bibnamefont
  {Cirac}},\ }\href {\doibase 10.1103/PhysRevB.84.165139} {\bibfield  {journal}
  {\bibinfo  {journal} {Phys. Rev. B}\ }\textbf {\bibinfo {volume} {84}},\
  \bibinfo {pages} {165139} (\bibinfo {year} {2011})}\BibitemShut {NoStop}%
\bibitem [{\citenamefont {Fidkowski}\ and\ \citenamefont
  {Kitaev}(2011)}]{LukaszfSPT1}%
  \BibitemOpen
  \bibfield  {author} {\bibinfo {author} {\bibfnamefont {L.}~\bibnamefont
  {Fidkowski}}\ and\ \bibinfo {author} {\bibfnamefont {A.}~\bibnamefont
  {Kitaev}},\ }\href {\doibase 10.1103/PhysRevB.83.075103} {\bibfield
  {journal} {\bibinfo  {journal} {Phys. Rev. B}\ }\textbf {\bibinfo {volume}
  {83}},\ \bibinfo {pages} {075103} (\bibinfo {year} {2011})}\BibitemShut
  {NoStop}%
\bibitem [{\citenamefont {Fidkowski}\ and\ \citenamefont
  {Kitaev}(2010)}]{LukaszfSPT2}%
  \BibitemOpen
  \bibfield  {author} {\bibinfo {author} {\bibfnamefont {L.}~\bibnamefont
  {Fidkowski}}\ and\ \bibinfo {author} {\bibfnamefont {A.}~\bibnamefont
  {Kitaev}},\ }\href {\doibase 10.1103/PhysRevB.81.134509} {\bibfield
  {journal} {\bibinfo  {journal} {Phys. Rev. B}\ }\textbf {\bibinfo {volume}
  {81}},\ \bibinfo {pages} {134509} (\bibinfo {year} {2010})}\BibitemShut
  {NoStop}%
\bibitem [{\citenamefont {Chen}\ \emph {et~al.}(2012)\citenamefont {Chen},
  \citenamefont {Gu}, \citenamefont {Liu},\ and\ \citenamefont
  {Wen}}]{XieSPT5}%
  \BibitemOpen
  \bibfield  {author} {\bibinfo {author} {\bibfnamefont {X.}~\bibnamefont
  {Chen}}, \bibinfo {author} {\bibfnamefont {Z.-C.}\ \bibnamefont {Gu}},
  \bibinfo {author} {\bibfnamefont {Z.-X.}\ \bibnamefont {Liu}}, \ and\
  \bibinfo {author} {\bibfnamefont {X.-G.}\ \bibnamefont {Wen}},\ }\href@noop
  {} {\bibfield  {journal} {\bibinfo  {journal} {Science 338, 1604}\ }
  (\bibinfo {year} {2012})}\BibitemShut {NoStop}%
\bibitem [{\citenamefont {Levin}\ and\ \citenamefont {Stern}(2012)}]{SPTCS1}%
  \BibitemOpen
  \bibfield  {author} {\bibinfo {author} {\bibfnamefont {M.}~\bibnamefont
  {Levin}}\ and\ \bibinfo {author} {\bibfnamefont {A.}~\bibnamefont {Stern}},\
  }\href {\doibase 10.1103/PhysRevB.86.115131} {\bibfield  {journal} {\bibinfo
  {journal} {Phys. Rev. B}\ }\textbf {\bibinfo {volume} {86}},\ \bibinfo
  {pages} {115131} (\bibinfo {year} {2012})}\BibitemShut {NoStop}%
\bibitem [{\citenamefont {Lu}\ and\ \citenamefont {Vishwanath}(2012)}]{SPTCS2}%
  \BibitemOpen
  \bibfield  {author} {\bibinfo {author} {\bibfnamefont {Y.-M.}\ \bibnamefont
  {Lu}}\ and\ \bibinfo {author} {\bibfnamefont {A.}~\bibnamefont
  {Vishwanath}},\ }\href {\doibase 10.1103/PhysRevB.86.125119} {\bibfield
  {journal} {\bibinfo  {journal} {Phys. Rev. B}\ }\textbf {\bibinfo {volume}
  {86}},\ \bibinfo {pages} {125119} (\bibinfo {year} {2012})}\BibitemShut
  {NoStop}%
\bibitem [{\citenamefont {Cheng}\ and\ \citenamefont {Gu}(2013)}]{SPTCS3}%
  \BibitemOpen
  \bibfield  {author} {\bibinfo {author} {\bibfnamefont {M.}~\bibnamefont
  {Cheng}}\ and\ \bibinfo {author} {\bibfnamefont {Z.-C.}\ \bibnamefont {Gu}},\
  }\href@noop {} {\bibfield  {journal} {\bibinfo  {journal} {arXiv:1302.4803}\
  } (\bibinfo {year} {2013})}\BibitemShut {NoStop}%
\bibitem [{\citenamefont {Levin}\ and\ \citenamefont {Gu}(2012)}]{duality}%
  \BibitemOpen
  \bibfield  {author} {\bibinfo {author} {\bibfnamefont {M.}~\bibnamefont
  {Levin}}\ and\ \bibinfo {author} {\bibfnamefont {Z.-C.}\ \bibnamefont {Gu}},\
  }\href {\doibase 10.1103/PhysRevB.86.115109} {\bibfield  {journal} {\bibinfo
  {journal} {Phys. Rev. B}\ }\textbf {\bibinfo {volume} {86}},\ \bibinfo
  {pages} {115109} (\bibinfo {year} {2012})}\BibitemShut {NoStop}%
\bibitem [{\citenamefont {Gu}\ and\ \citenamefont {Wen}(2012)}]{Gusuper}%
  \BibitemOpen
  \bibfield  {author} {\bibinfo {author} {\bibfnamefont {Z.-C.}\ \bibnamefont
  {Gu}}\ and\ \bibinfo {author} {\bibfnamefont {X.-G.}\ \bibnamefont {Wen}},\
  }\href@noop {} {\bibfield  {journal} {\bibinfo  {journal} {arXiv:1201.2648}\
  } (\bibinfo {year} {2012})}\BibitemShut {NoStop}%
\bibitem [{\citenamefont {Ryu}\ and\ \citenamefont {Zhang}(2012)}]{edge2}%
  \BibitemOpen
  \bibfield  {author} {\bibinfo {author} {\bibfnamefont {S.}~\bibnamefont
  {Ryu}}\ and\ \bibinfo {author} {\bibfnamefont {S.-C.}\ \bibnamefont
  {Zhang}},\ }\href {\doibase 10.1103/PhysRevB.85.245132} {\bibfield  {journal}
  {\bibinfo  {journal} {Phys. Rev. B}\ }\textbf {\bibinfo {volume} {85}},\
  \bibinfo {pages} {245132} (\bibinfo {year} {2012})}\BibitemShut {NoStop}%
\bibitem [{\citenamefont {Qi}(2012)}]{edge1}%
  \BibitemOpen
  \bibfield  {author} {\bibinfo {author} {\bibfnamefont {X.-L.}\ \bibnamefont
  {Qi}},\ }\href@noop {} {\bibfield  {journal} {\bibinfo  {journal}
  {arXiv:1202.3983}\ } (\bibinfo {year} {2012})}\BibitemShut {NoStop}%
\bibitem [{\citenamefont {Yao}\ and\ \citenamefont {Ryu}(2012)}]{edge3}%
  \BibitemOpen
  \bibfield  {author} {\bibinfo {author} {\bibfnamefont {H.}~\bibnamefont
  {Yao}}\ and\ \bibinfo {author} {\bibfnamefont {S.}~\bibnamefont {Ryu}},\
  }\href@noop {} {\bibfield  {journal} {\bibinfo  {journal} {arXiv:1202.5805}\
  } (\bibinfo {year} {2012})}\BibitemShut {NoStop}%
\bibitem [{Note1()}]{Note1}%
  \BibitemOpen
  \bibinfo {note} {This assumption is justified since we can always choose a
  basis of fermion operators in which $S$ is diagonal.}\BibitemShut {Stop}%
\bibitem [{\citenamefont {Schnyder}\ \emph {et~al.}(2008)\citenamefont
  {Schnyder}, \citenamefont {Ryu}, \citenamefont {Furusaki},\ and\
  \citenamefont {Ludwig}}]{RyuSPT}%
  \BibitemOpen
  \bibfield  {author} {\bibinfo {author} {\bibfnamefont {A.}~\bibnamefont
  {Schnyder}}, \bibinfo {author} {\bibfnamefont {S.}~\bibnamefont {Ryu}},
  \bibinfo {author} {\bibfnamefont {A.}~\bibnamefont {Furusaki}}, \ and\
  \bibinfo {author} {\bibfnamefont {A.~W.~W.}\ \bibnamefont {Ludwig}},\ }\href
  {\doibase 10.1103/PhysRevB.78.195125} {\bibfield  {journal} {\bibinfo
  {journal} {Phys. Rev. B}\ }\textbf {\bibinfo {volume} {78}},\ \bibinfo
  {pages} {195125} (\bibinfo {year} {2008})}\BibitemShut {NoStop}%
\bibitem [{\citenamefont {Kitaev}(2009)}]{Kitaevperiod}%
  \BibitemOpen
  \bibfield  {author} {\bibinfo {author} {\bibfnamefont {A.~Y.}\ \bibnamefont
  {Kitaev}},\ }\href@noop {} {\bibfield  {journal} {\bibinfo  {journal} {AIP
  Conf. Proc.}\ }\textbf {\bibinfo {volume} {1134}},\ \bibinfo {pages} {22}
  (\bibinfo {year} {2009})}\BibitemShut {NoStop}%
\bibitem [{\citenamefont {Kitaev}(2006)}]{KitaevSC}%
  \BibitemOpen
  \bibfield  {author} {\bibinfo {author} {\bibfnamefont {A.~Y.}\ \bibnamefont
  {Kitaev}},\ }\href@noop {} {\bibfield  {journal} {\bibinfo  {journal} {Annals
  of Physics}\ }\textbf {\bibinfo {volume} {321}},\ \bibinfo {pages} {2}
  (\bibinfo {year} {2006})}\BibitemShut {NoStop}%
\bibitem [{\citenamefont {Haldane}(1995)}]{Haldane}%
  \BibitemOpen
  \bibfield  {author} {\bibinfo {author} {\bibfnamefont {F.}~\bibnamefont
  {Haldane}},\ }\href {\doibase 10.1103/PhysRevLett.74.2090} {\bibfield
  {journal} {\bibinfo  {journal} {Phys. Rev. Lett.}\ }\textbf {\bibinfo
  {volume} {74}},\ \bibinfo {pages} {2090} (\bibinfo {year}
  {1995})}\BibitemShut {NoStop}%
\bibitem [{Note2()}]{Note2}%
  \BibitemOpen
  \bibinfo {note} {Note that it is not important to us whether these terms are
  relevant or irrelevant in the renormalization group sense. The reason is that
  we are not interested in the \protect \emph {perturbative} stability of the
  edge, but rather whether it is stable to arbitrary perturbations that do not
  break the symmetry, explicitly or spontaneously.}\BibitemShut {Stop}%
\bibitem [{Note3()}]{Note3}%
  \BibitemOpen
  \bibinfo {note} {Here, a fermionic state is said to be ``short-range
  entangled'' if it can be transformed into an atomic insulator by a local
  unitary transformation -- a unitary operator generated from the time
  evolution of a local Hamiltonian over a finite time $t$.}\BibitemShut {Stop}%
\bibitem [{Note4()}]{Note4}%
  \BibitemOpen
  \bibinfo {note} {For a more detailed proof in a related bosonic system, see
  Ref. \protect \rev@citealp {duality}.}\BibitemShut {Stop}%
\bibitem [{\citenamefont {Levin}\ and\ \citenamefont
  {Wen}(2003)}]{LevinWenHop}%
  \BibitemOpen
  \bibfield  {author} {\bibinfo {author} {\bibfnamefont {M.}~\bibnamefont
  {Levin}}\ and\ \bibinfo {author} {\bibfnamefont {X.-G.}\ \bibnamefont
  {Wen}},\ }\href {\doibase 10.1103/PhysRevB.67.245316} {\bibfield  {journal}
  {\bibinfo  {journal} {Phys. Rev. B}\ }\textbf {\bibinfo {volume} {67}},\
  \bibinfo {pages} {245316} (\bibinfo {year} {2003})}\BibitemShut {NoStop}%
\end{thebibliography}%

\end{document}